\documentclass[twocolumn,journal]{IEEEtran}
\usepackage[T1]{fontenc}
\usepackage[latin9]{inputenc}
\usepackage{verbatim}
\usepackage{float}
\usepackage{amsmath}
\usepackage{amssymb}
\usepackage{graphicx}
\usepackage{wrapfig}
\usepackage[unicode=true,
 bookmarks=true,bookmarksnumbered=true,bookmarksopen=true,bookmarksopenlevel=1,
 breaklinks=false,pdfborder={0 0 1},backref=false,colorlinks=false]
 {hyperref}
\hypersetup{pdftitle={Your Title},
 pdfauthor={Your Name},
 pdfborderstyle=,pdfpagelayout=OneColumn,pdfnewwindow=true,pdfstartview=XYZ,plainpages=false}

\makeatletter

\providecommand{\tabularnewline}{\\}
\floatstyle{ruled}
\newfloat{algorithm}{tbp}{loa}
\providecommand{\algorithmname}{Algorithm}
\floatname{algorithm}{\protect\algorithmname}

\AtBeginDocument{
\begin{acronym}[BEREC]
\acro{KM}{Kuhn-Munkres}
\acro{PLS}{physical layer security}
\acro{SRmax}{secrecy rate maximization}
\acro{V2X}{vehicle-to-everything}
\acro{CUE}{cellular user equipment}
\acro{BS}{base station}
\acro{V2I}{vehicle-to-infrastructure}
\acro{VUE}{vehicle-to-vehicle user equipment}
\acro{V2V}{vehicle-to-vehicle}
\acro{RB}{resource block}
\acro{OFDM}{orthogonal frequency division multiplexing}
\acro{SINR}{signal-to-interference-plus-noise ratio}
\acro{LOS}{line of sight}
\acro{NLOS}{non-line of sight}
\acro{SCA}{successive convex approximation}
\acro{A-FISTA}{alternating fast iterative shrinkage-thresholding algorithm}
\acro{AO}{alternating optimization}
\acro{VANET}{vehicular ad hoc network}
\acro{EVE}{eavesdropper}
\acro{OBU}{on-board unit}
\acro{DSRC}{dedicated short-range communications}
\acro{RSU}{roadside unit}
\acro{SOC}{second-order cone}

\acro{10G-EPON}{10 Gigabit EPON}
\acro{3GPP}{3rd Generation Partnership Project}
\acro{5G}{fifth generation}
\acro{6G}{sixth generation}
\acro{A-CPI}{A Controller Plane Interface}
\acro{A-RoF}{Analog Radio-over-Fiber}
\acro{ABNO}{Application-Based Network Operations}
\acro{ADC}{Analog-to-Digital Converter}
\acro{AE}{Allocative Efficiency}
\acro{AMC}{Adaptive Modulation and Coding}
\acro{AON}{Active Optical Network}
\acro{AI}{artificial intelligence}
\acro{ML}{machine learning}
\acro{API}{application programming interface}
\acro{AWG}{Arrayed Waveguide Grating}
\acro{B2B}{Business to Business}
\acro{B2C}{Business to Consumer}
\acro{B-RAS}{Broadband Remote Access Servers}
\acro{BaaS}{Blockchain as a Service}
\acro{BB}{Budget Balance}
\acro{BBF}{BroadBand Forum}
\acro{BBU}{Base Band Unit}
\acro{BER}{Bit Error Rate}
\acro{BEREC}{Body of European Regulators for Electronic Communications}
\acro{BF}{Beamforming}
\acro{BFT}{Byzantine Fault Tolerant}
\acro{BGP}{Border Gateway Protocol}
\acro{BMap}{Bandwidth Map}
\acro{BPM}{Business Process Management}
\acro{BS}{Base Station}
\acro{BSC}{Base Station Controller}
\acro{BtB}{Back to Back}
\acro{BW}{Bandwidth}
\acro{C2C}{Consumer to Consumer}
\acro{C-RAN}{Cloud Radio Access Network}
\acro{C-RoFN}{Cloud-based Radio over Optical Fiber Network}
\acro{CA}{Certificate Authority}
\acro{CapEx}{Capital Expenditure}
\acro{CDMA}{Code Division Multiple Access}
\acro{CDR}{Clock Data Recovery}
\acro{CFO}{Carrier Frequency Offset}
\acro{CIR}{Committed Information Rate}
\acro{CO}{Central Office}
\acro{CoMP}{Coordinated Multipoint}
\acro{COP}{Control Orchestration Protocol}
\acro{CORD}{Central Office Rearchitected as a Data Centre}
\acro{CPE}{Customer Premises Equipment}
\acro{CPRI}{Common Public Radio Interface}
\acro{CPU}{Central Processing Unit}
\acro{CMC}{Conflict Mitigation Controller}
\acro{CMS}{Conflict Management System}
\acro{CQI}{Channel Quality Indicator}
\acro{CR}{Cognitive Radio}
\acro{CS}{Central Station}
\acro{CSI}{channel state information}
\acro{D2D}{Device to Device}
\acro{D-CPI}{D Controller Plane Interface}
\acro{D-RoF}{Digital Radio-over-Fiber}
\acro{DAC}{Digital-to-Analog Converter}
\acro{DAS}{Distributed Antenna System}
\acro{DBA}{Dynamic Bandwidth Allocation}
\acro{DBRu}{Dynamic Bandwidth Report Upstream}
\acro{DC}{Direct Current}
\acro{DL}{Downlink}
\acro{DLT}{Distributed Ledger Technology}
\acro{DMT}{Discrete Multitone}
\acro{DNS}{Domain Name Server}
\acro{DPDK}{Data Plane Development Kit}
\acro{DSB}{Double Side Band}
\acro{DSL}{Digital Subscriber Line}
\acro{DSLAM}{Digital Subscriber Line Access Multiplexer}
\acro{DSP}{Digital Signal Processing}
\acro{DSS}{Distributed Synchronization Service}
\acro{DU}{Distributed Unit}
\acro{DUE}{D2D User Equipment}
\acro{DWDM}{Dense Wave Division Multiplexing}
\acro{E-CORD}{Enterprise CORD}
\acro{EAL}{Environment Abstraction Layer}
\acro{EDF}{Erbium-Doped Fiber}
\acro{EFM}{Ethernet in the First Mile}
\acro{EMBS}{Elastic Mobile Broadband Service}
\acro{EON}{Elastic Optical Network}
\acro{EPC}{Evolved Packet Core}
\acro{EPON}{Ethernet Passive Optical Network}
\acro{eMBB}{Enhanced Mobile Broadband}
\acro{ETSI}{European Telecommunications Standards Institute}
\acro{EVM}{Error Vector Magnitude}
\acro{FANS}{Fixed Access Network Sharing}
\acro{FASA}{Flexible Access System Architecture}
\acro{FBMC}{Filter Bank Multi-Carrier}
\acro{FCC}{Federal Communications Commission}
\acro{FDD}{Frequency Division Duplex}
\acro{FDMA}{Frequency-Division Multiple Access}
\acro{FEC}{Forward Error Correction}
\acro{FFR}{Fractional Frequency Reuse}
\acro{FFT}{Fast Fourier Transform}
\acro{FISTA}{Fast Iterative Shrinkage-Thresholding Algorithm}
\acro{FISTA-L}{FISTA with linesearch}
\acro{FPGA}{Field-Programmable Gate Array}
\acro{FSO}{Free-Space-Optics}
\acro{FSR}{Free Spectral Range}
\acro{FTN}{Faster-Than-Nyquist}
\acro{FTTC}{Fiber-to-the-Curb}
\acro{FTTH}{Fiber-to-the-Home}
\acro{FTTx}{Fiber-to-the-x}
\acro{G.Fast}{Fast Access to Subscriber Terminals}
\acro{GEM}{G-PON Encapsulation Method}
\acro{GFDM}{Generalized Frequency Division Multiplexing}
\acro{GMPLS}{Generalized Multi-Protocol Label Switching}
\acro{GPON}{Gigabit Passive Optical Network}
\acro{GPP}{General Purpose Processor}
\acro{GPRS}{General Packet Radio Service}
\acro{GSM}{Global System for Mobile Communications}
\acro{GTP}{GPRS Tunneling Protocol}
\acro{GUI}{Graphical User Interface}
\acro{HA}{Hardware Accelerator}
\acro{HARQ}{Hybrid-Automatic Repeat Request}
\acro{HD}{Half Duplex}
\acro{HetNet}{Heterogeneous Networks}
\acro{IaaS}{Infrastructure as a Service}
\acro{I-CPI}{I Controller Plane Interface}
\acro{I2RS}{Interface 2 Routing System}
\acro{IA}{Interference Alignment}
\acro{IAM}{Identity and Access Management}
\acro{IBFD}{In-Band Full Duplex}
\acro{IC}{Incentive Compatibility}
\acro{ICIC}{Inter-Cell Interference Coordination}
\acro{ICT}{Information and Communications Technology}
\acro{IEEE}{Institute of Electrical and Electronics Engineers}
\acro{IETF}{Internet Engineering Task Force}
\acro{IF}{Intermediate Frequency}
\acro{IFFT}{Inverse FFT}
\acro{ITS}{Intelligent Transport Systems}
\acro{InP}{Infrastructure Provider}
\acro{IoT}{Internet of Things}
\acro{IP}{Internet Protocol}
\acro{IQ}{In-Phase/Quadrature}
\acro{IR}{Individual Rationality}
\acro{IRC}{Interference Rejection Combining}
\acro{ISI}{Inter-Symbol Interference}
\acro{ISP}{Internet Service Provider}
\acro{IV}{Intelligent Vehicle}
\acro{ITU}{International Telecommunication Union}
\acro{KPI}{Key Performance Indicator}
\acro{KVM}{Kernel-based Virtual Machine}
\acro{L2}{Layer-2}
\acro{L3}{Layer-3}
\acro{LAN}{Local Area Network}
\acro{LO}{Local Oscillator}
\acro{LOS}{Line Of Sight}
\acro{LR-PON}{Long Reach PON}
\acro{LSP-DB}{Label Switched Path Database}
\acro{LTE-A}{Long Term Evolution Advanced}
\acro{LTE}{Long Term Evolution}
\acro{M-CORD}{Mobile CORD}
\acro{M-MIMO}{Massive MIMO}
\acro{M2M}{Machine-to-Machine}
\acro{MAC}{Medium Access Control}
\acro{ME}{Merging Engine}
\acro{MIMO}{Multiple Input Multiple Output}
\acro{MME}{Mobility Management Entity}
\acro{mmWave}{Millimeter Wave}
\acro{MNO}{Mobile Network Operator}
\acro{MPLS}{Multiprotocol Label Switching}
\acro{MRC}{Maximum Ratio Combining}
\acro{MSC}{Mobile Switching Centre}
\acro{MSP}{Membership Service Providers}
\acro{MSR}{Multi-Stratum Resources}
\acro{MTC}{Machine Type Communication}
\acro{mMTC}{Massive Machine Type Communications}
\acro{MVNO}{Mobile Virtual Network Operator}
\acro{MVNP}{Mobile Virtual Network Provider}
\acro{MZI}{Mach-Zehnder Interferometer}
\acro{MZM}{Mach-Zehnder Modulator}
\acro{NE}{Network Element}
\acro{NFV}{Network Function Virtualization}
\acro{NFVaaS}{Network Function Virtualization as a Service}
\acro{NG-PON2}{Next-Generation Passive Optical Network 2}
\acro{NGA}{Next Generation Access}
\acro{NLOS}{Non Line Of Sight}
\acro{NMS}{Network Management System}
\acro{NRZ}{Non Return-to-Zero}
\acro{NTT}{Nippon Telegraph and Telephone}
\acro{OBPF}{Optical BandPass Filter}
\acro{OBSAI}{Open Base Station Architecture Initiative}
\acro{ODN}{Optical Distribution Network}
\acro{OFC}{Optical Frequency Comb}
\acro{OFDM}{Orthogonal Frequency-Division Multiplexing}
\acro{OFDMA}{Orthogonal Frequency-Division Multiple Access}
\acro{OLO}{Other Licensed Operator}
\acro{OLT}{Optical Line Terminal}
\acro{ONF}{Open Networking Foundation}
\acro{ONU}{Optical Network Unit}
\acro{OOB}{Out-of-Band}
\acro{OOK}{On-Off Keying}
\acro{OpenCord}{Central Office Re-Architected as a Data Center}
\acro{OpEx}{Operating Expenditure}
\acro{OSI}{Open Systems Interconnection}
\acro{OSS}{Operations Support Systems}
\acro{OTT}{Over-the-Top}
\acro{OXM}{OpenFlow Extensible Match}
\acro{P2MP}{Ethernet over Point-to-Multipoint}
\acro{P2P}{Point-to-Point}
\acro{PAM}{Pulse Amplitude Modulation}
\acro{PAPR}{Peak-to-Average Power Rating}
\acro{PAPR}{Peak-to-Average Power Ratio}
\acro{PBMA}{Priority Based Merging Algorithm}
\acro{PCE}{Path Computation Elements}
\acro{PCEP}{Path Computation Element Protocol}
\acro{PCF}{Photonic Crystal Fiber}
\acro{PD}{Photodiode}
\acro{PDCP}{RD Control Protocol}
\acro{PDF}{Probability Distribution Function}
\acro{PGW}{Packet Gateway}
\acro{PHY}{Physical Layer}
\acro{PIR}{Peak Information Rate}
\acro{PMD}{Polarization Division Multiplexing}
\acro{PON}{Passive Optical Network}
\acro{POTS}{Plain Old Telephone Service}
\acro{PoET}{Proof of Elapsed Time}
\acro{PoW}{Proof of Work}
\acro{PoS}{Proof of Stake}
\acro{PTP}{Precision Time Protocol}
\acro{PWM}{Pulse Width Modulation}
\acro{QAM}{Quadrature Amplitude Modulation}
\acro{QoE}{Quality of Experience}
\acro{QoS}{Quality of Service}
\acro{QPSK}{Quadrature Phase Shift Keying}
\acro{R-CORD}{Residential CORD}
\acro{RA}{Resource Allocation}
\acro{RAN}{Radio Access Network}
\acro{RAT}{Radio Access Technology}
\acro{RFIC}{Radio Frequency Integrated Circuit}
\acro{RIC}{RAN Intelligent Controller}
\acro{RLC}{Radio Link Control}
\acro{RN}{Remote Node}
\acro{ROADM}{Reconfigurable Optical Add Drop Multiplexer}
\acro{RoF}{Radio-over-Fiber}
\acro{RRC}{Radio Resource Control}
\acro{RRH}{Remote Radio Head}
\acro{RRPH}{Remote Radio and PHY Head}
\acro{RRU}{Remote Radio Unit}
\acro{RSOA}{Reflective Semiconductor Optical Amplifier}
\acro{RU}{Remote Unit}
\acro{Rx}{Receiver}
\acro{SC-FDMA}{Single Carrier Frequency Division Multiple Access}
\acro{SCM}{Single Carrier Modulation}
\acro{SD-RAN}{Software Defined Radio Access Network}
\acro{SDAN}{Software Defined Access Network}
\acro{SDMA}{Semi-Distributed Mobility Anchoring}
\acro{SDN}{Software Defined Networking}
\acro{SDR}{Software Defined Radio}
\acro{SFBD}{Single Fiber Bi-Direction}
\acro{SGW}{Serving Gateway}
\acro{SI}{Self-Interference}
\acro{SIC}{Self-Interference Cancellation}
\acro{SIMO}{Single Input Multiple Output}
\acro{SLA}{Service Level Agreement}
\acro{SMF}{Single Mode Fiber}
\acro{SNMP}{Simple Network Management Protocol}
\acro{SNR}{Signal-to-Noise Ratio}
\acro{SP}{Service Providers}
\acro{Split-PHY}{Split Physical Layer}
\acro{SRI-OV}{Single Root Input/Output Virtualization}
\acro{SU}{Secondary User}
\acro{SUT}{System Under Test}
\acro{T-CONT}{Transmission Container}
\acro{TC}{Transmission Convergence}
\acro{TCO}{Total Cost of Ownership}
\acro{TD-LTE}{Time Division LTE}
\acro{TDD}{Time Division Duplex}
\acro{TDM}{Time Division Multiplexing}
\acro{TDMA}{Time Division Multiple Access}
\acro{TED}{Traffic Engineering Database}
\acro{TEID}{Tunnel Endpoint Identifier}
\acro{TLS}{Transport Layer Security}
\acro{TPS}{Transactions per Second}
\acro{TTI}{Transmission Time Interval}
\acro{TWDM}{Time and Wavelength Division Multiplexing}
\acro{Tx}{Transmitter}
\acro{UD-CRAN}{Ultra-Dense Cloud Radio Access Network}
\acro{UDP}{User Datagram Protocol}
\acro{UE}{User Equipment}
\acro{UF-OFDM}{Universally Filtered OFDM}
\acro{UFMC}{Universally Filtered Multicarrier}
\acro{UL}{Uplink}
\acro{UMTS}{Universal Mobile Telecommunications System}
\acro{URLLC}{Ultra-Reliable Low-Latency Communication}
\acro{USRP}{Universal Software Radio Peripheral}
\acro{vBBU}{Virtualized BBU}
\acro{vBMap}{Virtual Bandwidth Map}
\acro{vBS}{Virtual Base Station}
\acro{VCG}{Vickery-Clarke-Groves}
\acro{vCPE}{Virtual CPE}
\acro{vCPU}{Virtual Central Processing Unit}
\acro{vDBA}{Virtual DBA}
\acro{VDSL2}{Very-high-bit-rate Digital Subscriber Line 2}
\acro{VLAN}{Virtual Local Area Network}
\acro{VM}{Virtual Machine}
\acro{VNF}{Virtual Network Functions}
\acro{VNI}{Visual Networking Index}
\acro{VNO}{Virtual Network Operator}
\acro{VNTM}{Virtual Network Topology Manager}
\acro{vOLT}{Virtual OLT}
\acro{VPE}{Virtual Provider Edge}
\acro{VPN}{Virtual Private Network}
\acro{VTN}{Virtual Tenant Network}
\acro{VULA}{Virtual Unbundled Local Access}
\acro{WAN}{Wide Area Network}
\acro{WAP}{Wireless Access Point}
\acro{WCDMA}{Wide-Band Code Division Multiple Access}
\acro{WDM-PON}{Wavelength Division Multiplexing - Passive Optical Network}
\acro{WDM}{Wavelength Division Multiplexing}
\acro{WiMAX}{Worldwide Interoperability for Microwave Access}
\acro{WRPR}{Wired-to-RF Power Ratio}
\acro{XG-PON}{10 Gigabit PON}
\acro{XGEM}{XG-PON Encapsulation Method}
\acro{XOS}{XaaS Operating System}
\acro{Z}{Zulu}
\end{acronym} 
}
\usepackage{cite}
\usepackage{siunitx}

\usepackage[nolist]{acronym}

\usepackage{algorithmic}
\floatname{algorithm}{Algorithm}

\DeclareMathOperator{\maximize}{maximize}
\DeclareMathOperator{\minimize}{minimize}

\newcommand{\herm}{^{{\dagger}}}
\newcommand{\trans}{^{\mbox{\scriptsize T}}}

\usepackage[left=0.6in, right=0.6in, top=0.7in, bottom=1in,centering]{geometry}

\makeatother

\begin{document}
\title{\textcolor{black}{On the Sum Secrecy Rate Maximisation for Wireless Vehicular Networks}\vspace{-1mm}}
\author{Muhammad~Farooq,~\IEEEmembership{Member,~IEEE}, Le-Nam~Tran~\IEEEmembership{Senior~Member,~IEEE,}
Fatemeh~Golpayegani,~\IEEEmembership{Senior~Member,~IEEE}, and~Nima~Afraz,~\IEEEmembership{Senior~Member,~IEEE}\thanks{Muhammad~Farooq, Nima~Afraz, and Fatemeh~Golpayegani are with the
School of Computer Science, University College Dublin, Ireland, e-mail:
\protect\protect\protect\href{mailto:mfarooq@ieee.org}{mfarooq@ieee.org},\protect\protect\protect\href{mailto:nima.afraz@ucd.ie}{nima.afraz@ucd.ie},\protect\protect\protect\href{mailto:fatemeh.golpayegani@ucd.ie}{fatemeh.golpayegani@ucd.ie}.}\thanks{Le-Nam Tran is with the School of Electrical and Electronic Engineering,
University College Dublin, Ireland, e-mail: \protect\protect\protect\href{mailto:nam.tran@ucd.ie}{nam.tran@ucd.ie}.}\vspace{-8mm}}
\maketitle
\begin{abstract}
Wireless communications form the backbone of future vehicular networks,
playing a critical role in applications ranging from traffic control
to vehicular road safety. However, the dynamic structure of these
networks creates security vulnerabilities, making security considerations an integral part of network design. We address these security concerns from a physical layer security aspect by investigating achievable secrecy rates in wireless vehicular networks. Specifically, we aim to maximize the sum secrecy rate from all vehicular pairs subject to bandwidth and power resource constraints. For the considered problem, we first propose a solution based on the successive convex approximation (SCA) method, which has not been applied in this context before.  To further reduce the complexity of the SCA-based method, we also propose a low-complexity solution based on a fast iterative shrinkage-thresholding algorithm (FISTA). Our simulation results for SCA and FISTA show a trade-off between performance at convergence and runtime. While the SCA method achieves better performance at convergence, the FISTA-based approach is at least 300 times faster than the SCA method.
\end{abstract}

\begin{IEEEkeywords}
Resource allocation, vehicular network, successive convex approximation,
low complexity \vspace{-4mm}
\end{IEEEkeywords}

\section{Introduction}\vspace{-1mm}

\IEEEPARstart{W}{ireless} vehicular networks, comprising both cellular
users and automobiles have emerged as a significant component of the current
transportation systems, offering a wide range of applications \cite{Luo2020ITS}. Ensuring secure
communication inside these networks is of crucial importance to protect
sensitive information from \acp{EVE} \cite{Talpur2022MLITS}.
Several methods have been proposed to achieve security in wireless vehicular networks at different OSI layers. Among them, there is
a large body of studies
that look at \ac{PLS} approaches  \cite{Lee2016PowerAlloc,Lai2008Secrecy,Rodriguez2015PhySecurity,ElHalawany2019PhySecurity}, which is also the focus of this
paper.

An importation notion of \ac{PLS} is the so-called secrecy rate, which is the data rate at which a transmitter can send its information to an intended receiver without it being decoded by any \ac{EVE} \cite{Maurer93secrecy}. This perfect secrecy makes the transmitted information completely secure and immune to any form of attack, which can be achieved by a proper \textit{wiretap coding}  scheme \cite{Cai11serecy}. Thus, achievable \ac{SRmax} is naturally a critical topic in wireless vehicular networks. However, this faces numerous challenges due to the dynamic and convoluted nature of the network.
Furthermore, resource restrictions, such as limited bandwidth and
computational power, significantly add to the tractability of maximising secrecy rates while guaranteeing efficient network operation \cite{Barskar2016Secure}.
Also, coexisting with cellular users poses interference difficulties that
demand novel solutions to assure secure coexistence \cite{Lee2016PowerAlloc}.
Addressing these problems requires novel techniques to achieve
safe and efficient communication systems for vehicular networks.

To solve the issues discussed above, a limited number of studies are available in
the literature that considers a joint bandwidth resource and power management
for \ac{SRmax} in vehicular networks. The authors in
\cite{Yang2017VTC} presented a bisection search method to find power control in vehicular communication, but did not
consider bandwidth management. The authors in \cite{Liang2020DL}
provided a deep learning-based solution to the resource allocation
problem, but they also considered power management only, which was highly sensitive to channel variations too. 
In the context of joint bandwidth and power resource management, a 
notable work was \cite{liu2020secrecy} which considered the power
control along with the bandwidth reuse optimization. Specifically, the authors in
\cite{liu2020secrecy} used the \ac{KM} algorithm to solve
the \ac{SRmax} problem in cellular underlying \ac{V2V}
communication. However, the main drawback of the \ac{KM} algorithm is that
it solves the non-convex problem, which is computationally demanding
and may lead to inefficient solutions. In addition, it requires channel estimation at \ac{EVE} and possesses
a computational complexity of $\mathcal{O}(Z^4)$ (where $Z$ is a 3-D variable). Thus, for practical cases, we need a more accurate solution with lesser complexity.  

Against the above background, we consider the \ac{SRmax} in wireless
vehicular networks subject to the power constraint and the reuse coefficient
design constraint in this paper. Different from the related literature, we adopt
a \ac{SCA} framework to solve the formulated problem. 
Our choice is motivated by the success of applying this
optimization technique to similar settings, which was previously demonstrated
in \cite{Ngo2018CFMM}. 
Similar to \cite{liu2020secrecy}, we consider a \ac{VANET} model featuring simultaneous \ac{CUE} and \ac{VUE} pairs. To enhance resource utilization, the frequency spectrum is uniformly distributed into \acp{RB} allocated
to \acp{CUE}, which are subsequently reused by \ac{VUE} pairs.
While such reuse presents practical advantages, addressing the associated
challenges, such as interference management, remain a complex task
in the design of efficient \ac{VANET} systems. In this context,
our contributions are as follows: 
\begin{enumerate}
\item For the considered system model, we derive the achievable secrecy
rate for each \ac{VUE} pair in the corresponding \ac{RB}
by finding the \ac{SINR} expressions for each \ac{VUE} pair
and the eavesdropped channel. We formulate the sum \ac{SRmax}
problem for all \ac{VUE} pairs and all \acp{RB} in the presence
of power and bandwidth resource management constraints. 
\item  We propose a more efficient SCA method compared to existing solutions \cite{liu2020secrecy} which are based on the same framework. \ac{SCA} has been used before for similar
settings, so it is natural to adopt this method in our paper. While \ac{SCA} offers a general framework to deal with nonconvex problems, it is still
difficult to solve the considered problem as efficient convex approximations are required for large-scale settings. 
\item We also propose a more numerically efficient solution based on \ac{FISTA}
algorithm (which is a modified version of the proximal gradient method
\cite[Chapter 10]{beck2017first}). Since each convex subproblem in the SCA-based method 
is built on second-order methods, its complexity is relatively high. \textcolor{black}{Thus, the \ac{FISTA}-based method, which belongs to the class of first-order methods, has substantially
lower computational complexity than the \ac{SCA} method as indicated by the results. In contrast, the \ac{SCA} method offers better performance at convergence.}
\end{enumerate}
\vspace{-4mm}
\section{System Model}\vspace{-2mm}


We consider a scenario where \ac{CUE} and \ac{VUE} pairs are
served by a \ac{BS} in a single-cell \ac{VANET}. The \acp{CUE}
use \ac{OFDM} scheme to divide the system bandwidth $B$ among
the \acp{CUE} such that each \ac{CUE} uses one \ac{RB} of
$B/M$. During each \ac{RB}, \ac{VUE} pairs reuse the bandwidth
not being used by the corresponding \ac{CUE}.

\begin{figure}[tbh]
\vspace{-3mm}\centering\includegraphics[width=5.5cm]{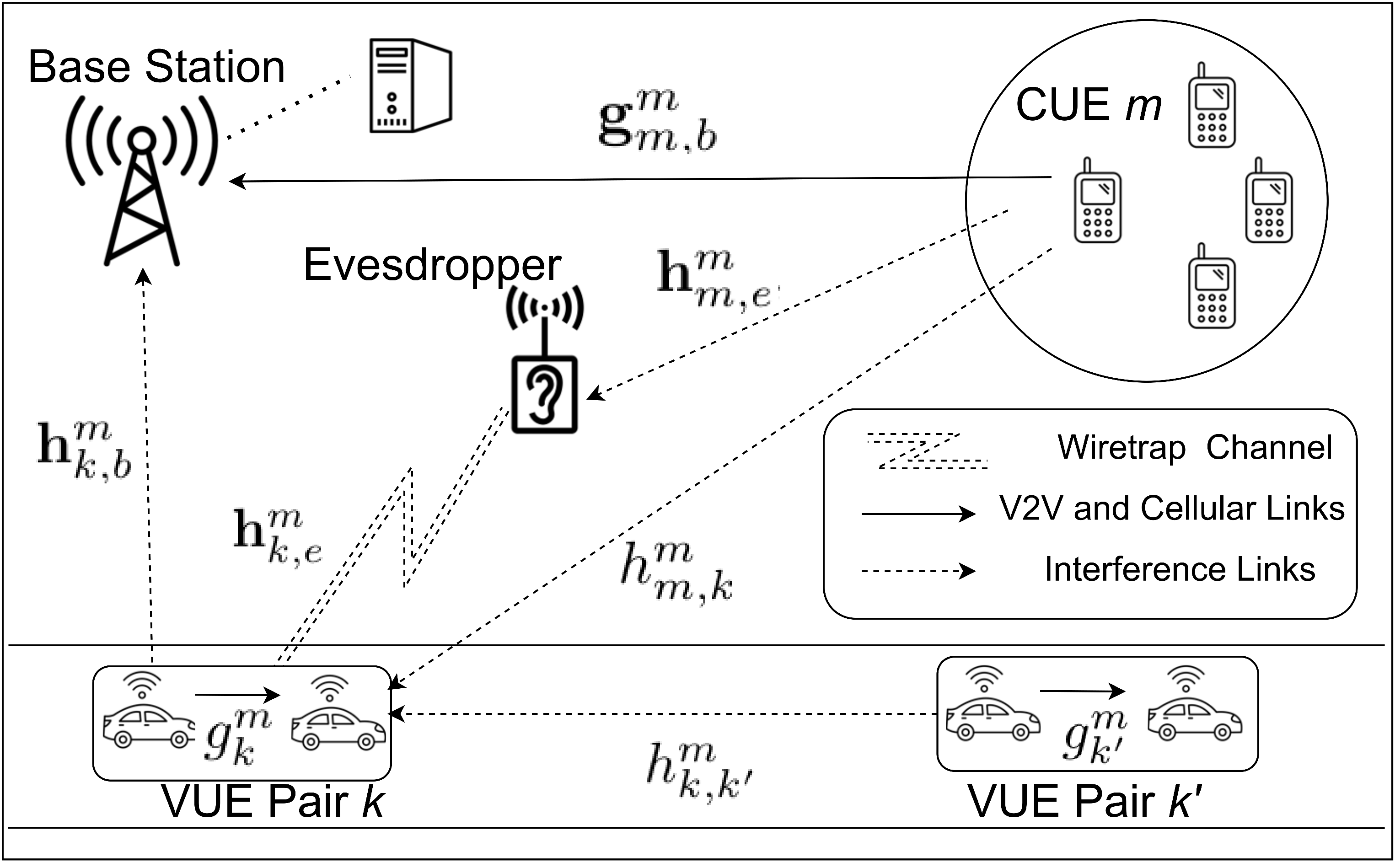}\vspace{-3mm}\caption{System Model}
\label{fig:systemmodel}\vspace{-3mm} 
\end{figure}

Consider the \ac{V2X} framework illustrated in Fig. \ref{fig:systemmodel}
which is a modified form of the system model considered in \cite{liu2020secrecy}.
In this setup, there are $M$ \acp{CUE} connected to \acp{BS}
and $K$ \ac{VUE} pairs that communicate with each other using
\ac{V2V} links. We represent these sets as $\mathcal{M}=\{1,2,\ldots,M\}$
and $\mathcal{K}=\{1,2,\ldots,K\}$, respectively. Each \ac{CUE}
and \ac{VUE} is equipped with one antenna. The \ac{BS} and the \ac{EVE}
are equipped with $N_{t}$ and $N_{e}$ antennas, respectively. 

The considered system is based on \ac{DSRC} technology \cite{Cseh1998DSRC}.
We divide the spectrum resources into $M$ \acp{RB}, each with
the same bandwidth, utilizing \ac{OFDM} technology. Initially,
each \ac{CUE} is assigned a specific \ac{RB} for uplink communication
with the \ac{BS}. However, \ac{VUE} pairs can efficiently reuse
the \acp{RB} allocated to \acp{CUE}, enhancing the overall spectrum
utilization. The total bandwidth, denoted as $B$, is partitioned
into $M$ \acp{RB}, with each \ac{CUE} $m$ occupying \ac{RB}
$m$. A binary index $q_{k}^{m}$ is employed to signify whether \ac{VUE}
pair $k$ reuses \ac{RB} $m$, with $q_{k}^{m}=1$ indicating reuse
and $q_{k}^{m}=0$ denoting otherwise.


As depicted in Fig. \ref{fig:systemmodel}, the channel gain between
\ac{CUE} $m$ and the \ac{BS} within a coherent time is characterized
by $\textcolor{black}{\mathbf{g}_{m,b}^{m}}\in\mathbb{C}^{N_{t}\times1}$. Additionally,
the channel gain $g_{k}^{m}$ represents the channel between \ac{VUE}
pair $k$ and \ac{RB} $m$. Given the \ac{RB} reuse, it is essential
to account for interference. Therefore, we define the interference
channel gain from \ac{VUE} transmitter $k$ to the \ac{BS} over
\ac{RB} $m$ as $\textcolor{black}{\mathbf{h}_{k,b}^{m}}\in\mathbb{C}^{N_{t}\times1}$
and the interference channel gain from \ac{CUE} $m$ to \ac{VUE}
receiver $k$ over \ac{RB} m as $h_{m,k}^{m}$. The interference
between \ac{VUE} pair $k$ and \ac{VUE} pair $k',k'\neq k$
is denoted as $h_{k,k'}^{m}$. The \ac{EVE} creates two channels,
including an interference channel between \ac{CUE} $m$ and \ac{EVE},
defined by an $N_{e}\times1$ vector, i.e., $\textcolor{black}{\mathbf{h}_{m,e}^{m}}$,
and a wiretap channel between \ac{VUE} $k$ and \ac{EVE}, defined
by a $N_{e}\times1$ vector, i.e., $\textcolor{black}{\mathbf{h}_{k,e}^{m}}$. \textcolor{black}{We assume that \ac{EVE} is a stationary listener and there is a specific time interval between two \ac{VUE} pairs passing through the same point. Thus, \ac{EVE} can interfere with only one \ac{VUE} pair in one \ac{RB}.}

\vspace{-3mm}
\section{Problem Formulation}
\vspace{-1mm}
We are interested in the achievable secrecy rate by the \ac{VUE}
pair $k$. For this purpose, we first write the received signal by
the \ac{VUE} pair $k$ which is given by 
\begin{equation}
y_{k}^{m}\!\!=\!\!q_{k}^{m}p_{k}^{m}g_{k}^{m}v_{k}^{m}\!\!+\!\!\textstyle\sum_{k'\neq k}q_{k'}^{m}p_{k'}^{m}h_{k,k'}^{m}v_{k'}^{m}\!+\!p_{m}^{m}h_{m,k}^{m}u_{m}^{m}\!+\!n_{k}^{m},
\end{equation}
where $q_{k}^{m}\in\{0,1\}$, $p_{k}^{m}$, and $v_{k}^{m}$ are the
reuse coefficient, the transmit power, and the signal communicated
by the \ac{VUE} pair $k$ in the \ac{RB} $m$, respectively.
\textcolor{black}{Similarly, $p_{m}^{m}$ and $u_{m}^{m}$ are the transmit power and the transmitted signal from the \ac{CUE} $m$
in the \ac{RB} $m$, respectively.} The signals $v_{k}^{m},\forall k$
and $u_{m}^{m},\forall m$ \textcolor{black}{are assumed to be i.i.d.} The variable $n_{k}^{m}\in\mathcal{CN}(0,\sigma^{2})$
is the noise signal. The received \ac{SINR} at the \ac{VUE}
pair $k$ over \ac{RB} m is expressed as 
\begin{equation}
\mathcal{S}_{k}^{m}=\tfrac{q_{k}^{m}p_{k}^{m}|g_{k}^{m}|^{2}}{\textstyle\sum_{k'\neq k}q_{k'}^{m}p_{k'}^{m}|h_{k,k'}^{m}|^{2}+p_{m}^{m}|h_{m,k}^{m}|^{2}+1}.\label{eq:SINRforVUE}
\end{equation}

The signal received by the \ac{EVE} eavesdropped from the \ac{VUE}
pair $k$ is given by 
\begin{equation}
\mathbf{y}_{e}^{m}=q_{k}^{m}p_{k}^{m}\mathbf{h}_{k,e}^{m}v_{k}^{m}+p_{m}^{m}\mathbf{h}_{m,e}^{m}u_{m}^{m}+\mathbf{n}_{e}^{m},
\end{equation}
where $\mathbf{n}_{e}^{m}\in\mathcal{CN}(0,\mathbf{I}_{N_{e}})$ is
a noise vector whose elements are i.i.d.. The \ac{EVE} used vector
$\mathbf{w}_{e}^{m}$ to decode the message from the \ac{VUE} pair
$k$, i.e., 
\begin{align}
 \!\!\!\!& \hat{y}_{e}^{m}=(\mathbf{w}_{k,e}^{m})\herm\mathbf{y}_{e}^{m}\nonumber \\
 \!\!\!\!& \!\!=\!\!q_{k}^{m}p_{k}^{m}(\mathbf{w}_{k,e}^{m})\herm\mathbf{h}_{k,e}^{m}v_{k}^{m}\!\!+\!p_{m}^{m}(\mathbf{w}_{k,e}^{m})\herm\mathbf{h}_{m,e}^{m}u_{m}^{m}\!\!+\!(\mathbf{w}_{k,e}^{m})\herm\mathbf{n}_{e}^{m}.
\end{align}
The optimal $\mathbf{w}_{k,e}^{m}$ which maximizes the eavesdropped
information is the eigenvector of the largest eigenvalue of $(p_{m}^{m}\mathbf{h}_{m,e}^{m}(\mathbf{h}_{m,e}^{m})\herm+\mathbf{n}_{e}^{m}(\mathbf{n}_{e}^{m})\herm)^{-1}\mathbf{h}_{k,e}^{m}(\mathbf{h}_{k,e}^{m})\herm$
\cite[Section II-B]{liu2020secrecy}. The \ac{SINR} for \ac{EVE}
is represented as
\begin{equation}
\textcolor{black}{\mathcal{S}_{k,e}^{m}=\tfrac{q_{k}^{m}p_{k}^{m}|(\mathbf{w}_{k,e}^{m})\herm\mathbf{h}_{k,e}^{m}|^{2}}{p_{m}^{m}|(\mathbf{w}_{k,e}^{m})\herm\mathbf{h}_{m,e}^{m}|^{2}+1}.}\label{eq:SINRforEVE}
\end{equation}
The capacity between \ac{VUE} pair $k$ over \ac{RB} $m$ is
expressed as
\begin{equation}
\textcolor{black}{\mathcal{C}_{k}^{m}=W\log_{2}(1+\mathcal{S}_{k}^{m}),}
\end{equation}
where $W$ is the bandwidth of an \ac{RB}. The capacity achieved
by the \ac{EVE} is given as 
\begin{equation}
\textcolor{black}{\mathcal{C}_{k,e}^{m}=W\log_{2}(1+\mathcal{S}_{k,e}^{m}),}
\end{equation}
Thus, in the presence of an \ac{EVE}, the achievable secrecy of
the \ac{VUE} pair $k$ is given by
\begin{equation}
R_{k}^{m}=max(\mathcal{C}_{k}^{m}-\mathcal{C}_{k,e}^{m}]_{+},0).
\end{equation}


We aim to maximize the sum secrecy rate of all \ac{VUE} pairs in
the presence of reuse coefficients constraint and \ac{V2V} power
constraint, which can be mathematically expressed as 

\begin{subequations}
\label{prob:main} 
\begin{align}
\underset{\mathbf{q},\mathbf{p}}{\maximize}\quad & \mathcal{R}(\mathbf{q},\mathbf{p})=\textstyle\sum_{m}\textstyle\sum_{k}R_{k}^{m}(\mathbf{q},\mathbf{p})\\
 & q_{k}^{m}\in\{0,1\},\forall k,\forall m,\label{eq:qbarconst}\\
 & 0<p_{k}^{m}\leq p_{\text{max}}^{m},\forall k,\forall m.\label{eq:pkmconst}
\end{align}
\end{subequations}
 Here $\mathbf{q}=[(\mathbf{q}^{1})\trans,(\mathbf{q}^{2})\trans,\ldots,(\mathbf{q}^{M})\trans],\trans$,
where $(\mathbf{q}^{m})\trans\triangleq[q_{1}^{m},\ldots,q_{K}^{m}]$
and $\mathbf{p}\triangleq[(\mathbf{p}^{1})\trans,(\mathbf{p}^{2})\trans,\ldots,(\mathbf{p}^{M})\trans]$,
where $(\mathbf{p}^{m})\trans\triangleq[p_{1}^{m},\ldots,p_{K}^{m}]$.
\vspace{-3mm}
\section{Proposed Solution }\vspace{-2mm}
In this section, we propose two methods for solving \eqref{prob:main}.
First, we provide a solution based on \ac{SCA} method and later
propose an \ac{FISTA} algorithm for solving the formulated problem.
\vspace{-4mm}
\subsection{Problem Simplification}
\vspace{-1mm}
Before presenting the proposed solutions, we make two important observations, which indeed make the considered problem easier to solve.
First, we note that the variables $q_{k}^{m}$ and $p_{k}^{m},\forall k,m$
are coupled in \ac{SINR} expressions in \eqref{eq:SINRforVUE} and
\eqref{eq:SINRforEVE}. However, we also note that when $q_{k}^{m}=0,$
\ac{SINR} becomes zero. Thus, we must only consider the case where
$q_{k}^{m}=1$ where $p_{k}^{m}$ becomes the only optimization variable.
We propose a way to reformulate the \ac{SINR} expressions by combining
$q_{k}^{m}$ and $p_{k}^{m}$ into one variable $p_{k}^{m}$, i.e.,
$q_{k}^{m}p_{k}^{m}=p_{k}^{m}$ such that when $q_{k}^{m}=0$, $p_{k}^{m}$
and hence \ac{SINR} become zero. The constraints \eqref{eq:qbarconst}
and \eqref{eq:pkmconst} can thus be combined into one constraint
\begin{equation}
0<p_{k}^{m}\leq p_{\text{max}}^{m}q_{k}^{m},\forall k,\forall m,\text{ where }q_{k}^{m}\in\{0,1\}.\label{eq:jointpq}
\end{equation}

Secondly, in practical wireless networks, there are several techniques
deployed to improve to degrade the strength of \ac{EVE} \cite{Shiu2011Security}
making it nearly impossible to get a zero secrecy rate. In other words,
we can say that $q_{k}^{m}=0$ is a highly unlikely case. Thus, with
these two observations, the problem in \eqref{prob:main} can be simplified
into \eqref{prob:main-2}.
\begin{subequations}
\label{prob:main-2} 
\begin{align}
\underset{\mathbf{p}}{\maximize}\quad & \mathcal{R}(\mathbf{p})=\textstyle\sum_{m}\textstyle\sum_{k}R_{k}^{m}(\mathbf{p})\\ 
 & 0<p_{k}^{m}\leq p_{\text{max}}^{m},\forall k,\forall m.
\end{align}
\end{subequations}
In the following subsections, we present the proposed solutions to
the simplified problem in \eqref{prob:main-2}.
\vspace{-3mm}
\subsection{SCA Based Solution}
\vspace{-1mm}
We are now in a position to propose the \ac{SCA} algorithm. In \ac{SCA},
we use a series of approximations to convert the non-convex problem
into a convex program. This has an advantage over the existing KM method, \ac{SCA} is proven to find a locally optimal solution \cite{Ngo2018CFMM}. We know that the problem in
\eqref{prob:main-2} can be approximated by \eqref{prob:mainapprox}.
\begin{subequations}
\label{prob:mainapprox} 
\begin{align}
\underset{\mathbf{x}}{\maximize}\quad & z=\textstyle\sum_{m}\textstyle\sum_{k}[\zeta_{k}^{m}-\gamma_{k}^{m}]_{+}\\
 & 0<p_{k}^{m}\leq p_{\text{max}}^{m},\forall k,\forall m,\label{eq:pconstraint}\\
 & C_{k}^{m}\geq\zeta_{k}^{m},\forall k,\forall m\label{Ckconst}\\
 & C_{ke}^{m}\leq\gamma_{k}^{m},\forall k,\forall m\label{Ckeconst}\\
 & \zeta_{k}^{m}\geq0,\gamma_{k}^{m}\geq0,\forall k,\forall m,\label{eq:zetagammaconst}
\end{align}
\end{subequations}
 where $\zeta_{k}^{m}$ and $\gamma_{k}^{m}$ are slack variables, and $\mathbf{x}\triangleq\big\{\{p_{k}^{m}\},\{\zeta_{k}^{m}\},\{\gamma_{k}^{m}\}\big\}.$
The only troublesome constraints in the above problem are \eqref{Ckconst}
and \eqref{Ckeconst} which can be written in the form of \ac{SOC}
as given in \eqref{eq:soc1} and \eqref{eq:soc2}. 
\begin{subequations}
\begin{equation}
2^{\zeta_{k}^{m}\!-\!1}\Big[\textstyle\sum_{k'\neq k}p_{k'}^{m}|h_{k,k'}^{m}|^{2}\!+\!p_{m}^{m}|h_{m,k}^{m}|^{2}\!+\!1\Big]\leq p_{k}^{m}|g_{k}^{m}|^{2},\label{eq:soc1}
\end{equation}
\begin{equation}
p_{k}^{m}|(\mathbf{w}_{k,e}^{m})\herm\mathbf{h}_{k,e}^{m}|^{2}\!\leq\!2^{\gamma_{k}^{m}\!-\!1}\!\Big[p_{m}^{m}|(\mathbf{w}_{e}^{m})\herm\mathbf{h}_{m,e}^{m}|^{2}\!\!+\!\!1\!\Big].\label{eq:soc2}
\end{equation}
\end{subequations}
 We need to find convex approximations of the constraints \eqref{eq:soc1}
and \eqref{eq:soc2}. Let $x_{1}=2^{\zeta_{k}^{m}-1}$, $y_{1}=\textstyle\sum_{k'\neq k}p_{k'}^{m}|h_{k,k'}^{m}|^{2}+p_{m}^{m}|h_{m,k}^{m}|^{2}+1$,
$x_{2}=2^{\gamma_{k}^{m}-1},$ and $y_{2}=p_{m}^{m}|(\mathbf{w}_{e}^{m})\herm\mathbf{h}_{m,e}^{m}|^{2}+1$
then these constraints become $E_{1}:x_{1}y_{1}\leq t_{1}$ and $E_{2}:x_{2}y_{2}\geq t_{2}$,
where $t_{1}=p_{k}^{m}|g_{k}^{m}|^{2}$ and $p_{k}^{m}|\!(\mathbf{w}_{k,e}^{m})\herm\mathbf{h}_{k,e}^{m}|^{2}$.
We recall that $xy=\tfrac{1}{4}[(x+y)^{2}-(x-y)^{2}]$. It can be
noted that we need a convex upper bound of $xy$ for the convex approximations
of $E_{1}$ and a concave lower bound of $xy$ for the convex approximation
of $E_{2}$, which are respectively expressed as \cite{Farooq2023FL}
\begin{subequations}
\begin{align}
\negthickspace\negthickspace x_{1}y_{1}\negthickspace & \leq\negthickspace\tfrac{1}{4}[(x_{1}\negthickspace+\negthinspace y_{1})^{2}\negthickspace-\negthickspace2(x_{1}\negthickspace-\negthickspace y_{1})(x_{1}^{n}\negthickspace-\negthickspace y_{1}^{n})\negthickspace+\negthickspace(x_{1}^{n}\negthickspace-\negthickspace y_{1}^{n})^{2}]\negthickspace\leq\negthickspace t_{1},\label{eq:soc1capprox}\\
\negthickspace\negthickspace x_{2}y_{2}\negthickspace & \geq\negthickspace\tfrac{1}{4}[2(x_{2}\negthickspace+\negthickspace y_{2})(x_{2}^{n}\negthickspace+\negthickspace y_{2}^{n})\negthickspace-\negthickspace(x_{2}^{n}\negthickspace+\negthickspace y_{2}^{n})^{2}\negthickspace-\negthickspace(x_{2}\negthickspace-\negthickspace y_{2})^{2}]\negthickspace\geq\negthickspace t_{2}.\label{eq:soc2capprox}
\end{align}
\end{subequations}
 Each iteration of the \ac{SCA} algorithm is approximated by the
following convex problem: 
\begin{align}
\maximize & \quad\{z\,|\,\mathbf{x}\in\mathcal{F}\},\label{eq:convexproblem}
\end{align}
where $\mathcal{F}\triangleq\{\eqref{eq:pconstraint},\eqref{eq:zetagammaconst},\eqref{eq:soc1capprox},\eqref{eq:soc2capprox}\}$.
The \ac{SCA} algorithm description is outlined in Algorithm \ref{alg:SCA}.
\vspace{-2mm}
\begin{algorithm}[th!]
\caption{\ac{SCA} Algorithm for Solving \ref{prob:mainapprox}}
\label{alg:SCA}
\begin{algorithmic}[1]
\STATE Input: $\mathbf{x}^{(0)}$,$n=1$
\WHILE{ convergence}
\STATE $\mathbf{x}^{(n+1)}\leftarrow$ Solve \eqref{eq:convexproblem}
\STATE $n=n+1$
\ENDWHILE
\STATE Output: $\mathbf{x}^{*}$
\end{algorithmic} 
\end{algorithm}
\vspace{-5mm}

\subsection{Proposed FISTA Algorithm}
 Although methods based on the \ac{SCA} framework have proved to be very efficient, they generally require
 high complexity due to the use of interior point convex solvers such as MOSEK for our problem. 
To further reduce the complexity of the proposed SCA method, we now propose a second method based on the \ac{FISTA} algorithm to solve the
approximate problem \eqref{prob:main-2}, which is more computationally efficient. The proposed \ac{FISTA}-based algorithm is
stated in Algorithm \ref{alg:mAPG}. \textcolor{black}{Here, we define $\mathcal{P}=\{\mathbf{p}\,|\,0<p_{m}^{m}\leq p_{\text{max}}^{m},\forall m\in M\ \text{and}\ 0<p_{k}^{m}\leq p_{\text{max}}^{m},\forall k\in K,\forall m\in M\}$ and choose $\alpha_p<1/L$, where $L$ is a Lipschitz constant.}
\vspace{-3mm}
\begin{algorithm}[th!]
\caption{Proposed \ac{FISTA} Algorithm for Solving \ref{prob:mainapprox}}
\label{alg:mAPG}
\begin{algorithmic}[1]
\STATE Input: \textcolor{black}{$\mathbf{p}^{(0)}\in\mathcal{P}, \alpha_{p}<1/L$, $n=1$}
\WHILE{ convergence}
\STATE $\mathbf{p}^{(n)}=\Phi_{\mathcal{P}}(\mathbf{p}^{(n-1)}+\alpha_{p}\nabla_{\mathbf{p}}\mathcal{R}(\mathbf{p}^{(n-1)}))$
\STATE $n=n+1$
\ENDWHILE
\STATE Output: $\mathbf{p}^{*}$
\end{algorithmic} 
\end{algorithm}
\vspace{-5mm}

Each iteration of the Algorithm \ref{alg:mAPG} requires
mainly two components: gradient w.r.t. $\mathbf{p}$ and projection
onto $\mathcal{P}$. In the following, we provide closed-form expressions
for these components.

\subsubsection{Gradient w.r.t. \texorpdfstring{$\mathbf{p}$}{p}}
The gradient w.r.t. $\mathbf{p}$ is given as
\begin{equation}
\nabla_{\mathbf{p}}\mathcal{R}(\mathbf{p})\!\!=\!\!\textstyle\sum_{m}\!\textstyle\sum_{k}\!\nabla_{\mathbf{p}}R_{k}^{m}(\mathbf{p})\!\!=\!\!\textstyle\sum_{m}\!\textstyle\sum_{k}\!\nabla_{\mathbf{p}}[\mathcal{C}_{k}^{m}-\mathcal{C}_{k,e}^{m}]_{+}.
\end{equation}
We know that $\nabla_{\mathbf{p}}(X)=[(\nabla_{\mathbf{p}^{1}}(X))\trans,\ldots,(\nabla_{\mathbf{p}^{M}}(X))\trans]\trans$.
Further, $\nabla_{\mathbf{p}^{m}}=[(\nabla_{p_{1}^{m}}(X))\trans,\ldots,(\nabla_{p_{K}^{m}}(X))\trans]\trans$.
The gradients $\nabla_{p_{i}^{m}}\mathcal{C}_{k}^{m}$ and $\nabla_{p_{i}^{m}}\mathcal{C}_{k,e}^{m}$
are given as

\begin{equation}
\nabla_{p_{i}^{m}}\mathcal{C}_{k}^{m}=\begin{cases}
\tfrac{W}{A+B}(|g_{k}^{m}|^{2}), & i=k\\
\tfrac{W}{A+B}(|h_{k,i}^{m}|^{2})-\tfrac{W}{B}(|h_{k,i}^{m}|^{2}), & i\neq k
\end{cases},
\end{equation}
\begin{equation}
\nabla_{p_{i}^{m}}\mathcal{C}_{k,e}^{m}=\begin{cases}
\tfrac{W}{C+D}|(\mathbf{w}_{k,e}^{m})\herm\mathbf{h}_{k,e}^{m}|^{2}, & i=k\\
0, & i\neq k
\end{cases},
\end{equation}
where $A=p_{k}^{m}|g_{k}^{m}|^{2}$, $B=\textstyle\sum_{k'\neq k}p_{k'}^{m}|h_{k,k'}^{m}|^{2}+p_{m}^{m}|h_{m,k}^{m}|^{2}+\sigma^{2}$,
$C=p_{k}^{m}|(\mathbf{w}_{k,e}^{m})\herm\mathbf{h}_{k,e}^{m}|^{2}$,
and $D=p_{m}^{m}|(\mathbf{w}_{e}^{m})\herm\mathbf{h}_{k,e}^{m}|^{2}+|(\mathbf{w}_{k,e}^{m})\herm\mathbf{n}_{e}^{m}|^{2}$. 

\subsubsection{\label{sec:projection}Projection onto \texorpdfstring{$\mathcal{P}$}{P}}

For a given $\mathbf{x}$, projection onto $\mathcal{P}$ is the solution
to the following problem:
\begin{multline}
\Phi_{\mathcal{P}}(\mathbf{x})=\underset{\mathbf{p}}{\minimize}\ \Bigl\{\Vert\mathbf{p}-\mathbf{x}\Vert^{2}\ \Bigl|\ 0<\mathbf{p}\leq p_{max}^{m}\Bigr\}.
\end{multline}
It is straightforward to check that the solution to the above problem
is 
\begin{equation}
\mathbf{p}=\min(\mathbf{x},p_{\max}^{m}).
\end{equation}
\vspace{-3mm}

\textcolor{black}{In Algorithm \ref{alg:mAPG}, we need to select a stepsize such that
$\alpha_{p}<1/L$.} Practically,
finding the Lipshitz constant and hence the stepsize could be challenging.
Therefore, to tune the stepsize, we use linesearch in each iteration
of the FISTA algorithm, i.e., we find the largest stepsize which satisfies
\begin{equation}
\mathcal{R}(\mathbf{p})\!\!\leq\!\mathcal{R}(\mathbf{p}^{n-1})\!+\!\delta\Vert\Phi_{\mathcal{P}}(\mathbf{p}^{n-1}\!+\!\alpha_{p}\nabla_{\mathbf{p}}\mathcal{R}(\mathbf{p}^{n\!-\!1}))\!-\!\mathbf{p}^{n\!-\!1}\Vert^{2}.
\end{equation}
\textcolor{black}{We refer to the modified algorithm with linesearch as \ac{FISTA-L} in the results sections.}

\vspace{-4mm}
\subsection{\label{subsec:Complexity-Analysis}Complexity Analysis}
\vspace{-1mm}
The per-iteration computational complexity of the proposed \ac{SCA}
algorithm is $\mathcal{O}(M^{3.5}K^{3.5})$ \cite{Ngo2018CFMM}.
For the computational complexity of \ac{FISTA}, we note that finding
$\mathcal{R}(\mathbf{q},\mathbf{p})$ and $\nabla_{\mathbf{p}}\mathcal{R}(\mathbf{q},\mathbf{p})$
requires $\mathcal{O}(MK)$ multiplications. Projection onto $\mathcal{P}$  requires $\mathcal{O}(K)$ operations. Hence,
the per-iteration complexity of the \ac{FISTA} algorithm is $\mathcal{O}(MK)$.

\begin{figure*}
\begin{minipage}[t]{0.33\linewidth}%
\begin{center}
\includegraphics[width=5cm]{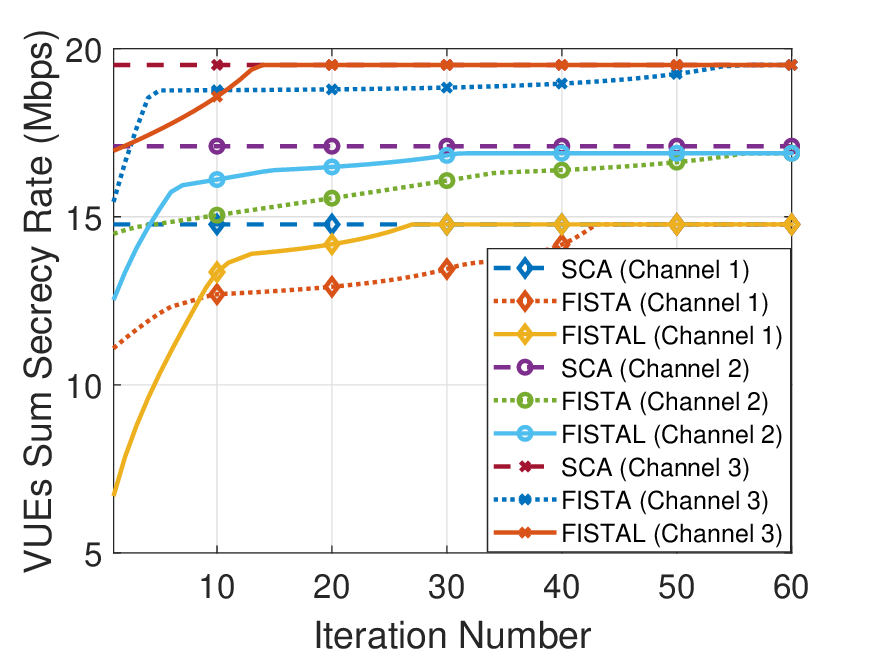}\vskip-0.1in\caption{$M=K=4$, $N_{t}=4$, and $N_{e}=2$.}
\label{fig:convergence} 
\par\end{center}%
\end{minipage}\hfill{}%
\begin{minipage}[t]{0.33\linewidth}%
\begin{center}
\includegraphics[width=5cm]{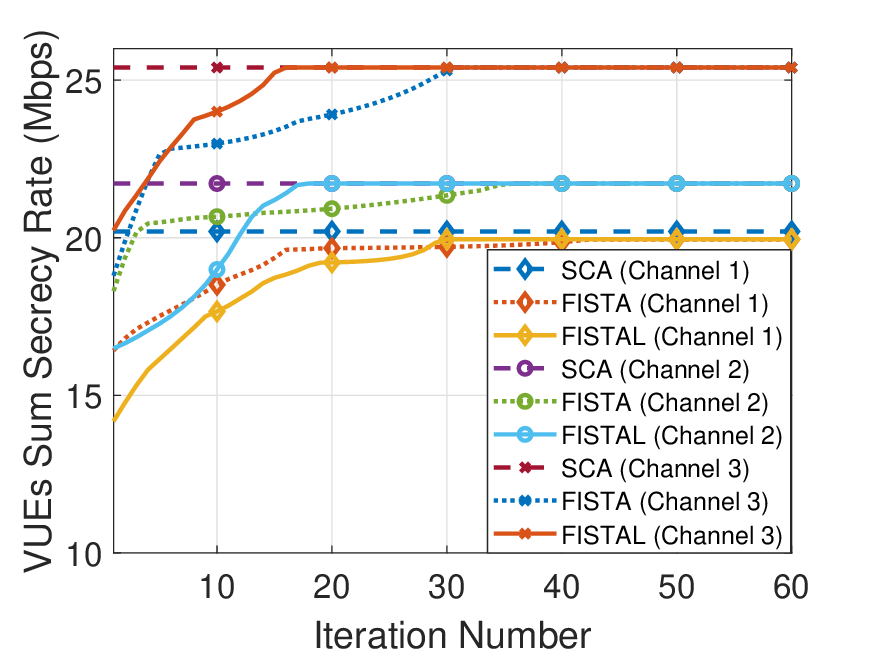}\vskip-0.1in\caption{$M=K=8$, $N_{t}=8$, and $N_{e}=4$.}
\label{fig:convergence-1}
\par\end{center}%
\end{minipage}\hfill{}%
\begin{minipage}[t]{0.33\linewidth}%
\begin{center}
\includegraphics[width=5cm]{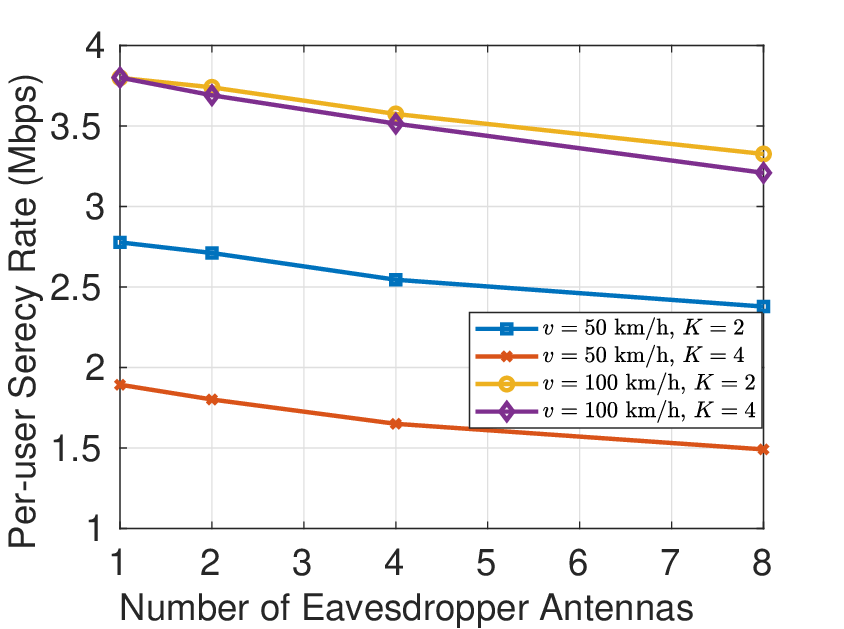}\vskip-0.1in\caption{$M=4$ and $N_{t}=4$.}
\label{fig:Ne-1} 
\par\end{center}%
\end{minipage}\vspace{-5mm}
\end{figure*}

\vspace{-3mm}
\section{Numerical Results}
\vspace{-1mm}
In \acp{VANET}, automobiles are equipped with \acp{OBU} that
permit communication either among themselves or with \acp{RSU}
utilising the \ac{DSRC} protocol \cite{camp2005vehicle}. As per
the \ac{DSRC} standard, a vehicle normally transmits signals within
a time interval ranging from 100 to 300 ms \cite{Zhang2008RSU}. The
effective communication range between vehicles is often between 50
m and 300 m, while the communication range of \acp{RSU} is typically
constrained to roughly 1000 m \cite{Willke2009}.

We assume a coherence time of 200 ms. Each vehicle transmits signals
within this coherence time. The communication range between \ac{V2V}
pair and between the vehicle and \ac{BS} are set to 100 m and
1000 m, respectively. During each coherence interval, vehicles on
urban area roads (maximum speed = 50 km/h) and national highways (maximum
speed = 100 km/h) can cover a distance of 1.39 m and 2.78 m, respectively.
Since the distance covered by the vehicles is substantially lower
than the coverage area (1000 m), we assume \ac{V2X} and \ac{V2V}
channels to remain constant during each coherence interval, which is
a reasonable assumption. Further, we assume each next vehicle passes
a certain point on the road after almost 5 seconds. Thus, when $v=50$
km/h, the distance between the neighbouring \ac{VUE} pairs will
be $\approx69.45$ m which is within the coverage area so inter-\ac{VUE}
interference will be present. For $v=100$ km/h, this distance becomes
$\approx138.9$ m which is outside the coverage area so inter-\ac{VUE}
interference can be neglected.

\vspace{-4mm}
\subsection{Simulation Parameters}
\vspace{-1mm}
If not otherwise mentioned, we set $p_{m}^{m}=1$ W, $p_{\max}^{m}=1$
W, $B=20$ MHz, $\delta=10^{-5}$, and $v=50$ km/h. Both iterative
methods are terminated when the difference of the objective for the
latest two iterations is less than $10^{-5}$. For the \ac{SCA}
method, we use convex conic solver MOSEK 
through
the modelling tool YALMIP. 
Each channel instance
in Fig. \ref{fig:systemmodel} is modelled as a Rician fading channel
\cite{liu2020secrecy}. In the context of Rician fading, the channel
response comprises a deterministic line-of-sight (LOS) component and
a stochastic non-line-of-sight (NLOS) component, as given below. 
\begin{equation}
h=\sqrt{\tfrac{1}{1+k}}(\sqrt{k}h_{\text{LOS}}+h_{\text{NLOS}}).
\end{equation}

\vspace{-5mm}
\subsection{Results}
\vspace{-1mm}
We remark that the considered system is different from \cite{liu2020secrecy} as they consider channel estimation at \ac{EVE} (which is practically insecure), and thus, has not been considered in this paper. Moreover, the computational complexity of the \ac{KM} method is $\mathcal{O}(Z^4)$, where $Z$ is a 3-D variable, making it difficult to run even for small-scale settings. However, \ac{SCA} and \ac{FISTA} methods have computational complexity of $\mathcal{O}(M^{3.5}K^{3.5})$ (including YALMIP's modelling complexity) and $\mathcal{O}(MK)$, respectively. Thus, we only provide numerical results for \ac{SCA} and \ac{FISTA} methods in this section.

We first compare the performance at convergence of the proposed \ac{FISTA} and
\ac{FISTA-L} methods against the solution from the proposed
\ac{SCA} method. In Figs. \ref{fig:convergence} and \ref{fig:convergence-1},
we consider two different sets of parameters: (i) $M=K=4$, $N_{t}=4$,
and $N_{e}=2$ and (ii) $M=K=8$, $N_{t}=8$, and $N_{e}=4$, and
three randomly generated channels for each of these sets of parameters.
It can be observed that for both scenarios, the \ac{FISTA} and \ac{FISTA-L}
methods nearly converge to the same solution as the solution
achieved using the \ac{SCA} method. For some channels (channel
2 in Fig. \eqref{fig:convergence} and channel 3 in Fig. \eqref{fig:convergence-1}),
we observe $1\%$ less objective from the first-order methods, which
makes \ac{SCA} superior in terms of performance.

The main advantage of the first order
methods over the \ac{SCA} method lie in the lower computational
complexity, as given in Section \eqref{subsec:Complexity-Analysis}. It can
be observed in Table \ref{tab:runtime} that, for three different
scenarios, the runtime of \ac{FISTA} and \ac{FISTA-L} methods
is significantly lower than the \ac{SCA} method. 

\vspace{-5mm}
\begin{table}[ht]
\centering \caption{Average runtime (in seconds) comparison between \ac{SCA} and first-order
(\ac{FISTA} and \ac{FISTA-L}) algorithms. Scenarios 1: $M=K=4$,
$N_{t}=4$, $N_{e}=2$; Scenarios 2: $M=K=6$, $N_{t}=6$, $N_{e}=3$;
Scenarios 3: $M=K=8$, $N_{t}=8$, $N_{e}=4$.}
\begin{tabular}{|c|c|c|c|}
\hline 
\textbf{Algorithm}  & \textbf{Scenario $1$}  & \textbf{Scenario $2$}  & \textbf{Scenario $3$}\tabularnewline
\hline 
\hline 
\ac{SCA}  & 115.92  & 412.53  & 509.10\tabularnewline
\hline 
\ac{FISTA}  & 0.39  & 0.85  & 1.48\tabularnewline
\hline 
\ac{FISTA-L}  & 0.05  & 0.12  & 0.2\tabularnewline
\hline 
\end{tabular}\label{tab:runtime} 
\vspace{-3mm}
\end{table}

\textcolor{black}{The choice between \ac{SCA} and \ac{FISTA} offers benefits, but each has its own drawbacks. \ac{SCA}'s accurate optimizations can boost network performance and security, though it takes its time, which might slow down real-time responses. On the other hand, \ac{FISTA} is fast and keeps the network adaptable, but it may not always achieve maximum performance or security. Both methods have their strengths, with just a few trade-offs along the way.}

Finally, we want to look at the aspect of how increasing the strength
of \ac{EVE} affects the achievable secrecy rate. For this purpose,
we plot the per-user secrecy rate against the number of \ac{EVE}
antennas ($N_{e}$) averaged over $100$ channel realisations as shown in Fig. \ref{fig:Ne-1}. As we increase
$N_{e}$, the per-user secrecy rate decreases, which is expected because
the \ac{EVE} gets more strength to eavesdrop on the desired channel.
In addition, when the number of \ac{VUE} pairs increases, the per-user
secrecy rate decreases due to inter-user interference. Moreover, the
decrease in the per-user is higher for low speeds ($v=50$ km/h) and
almost negligible for high speeds ($v=100$ km/h). This is due to
the reason that at $v=50$ km/h, the increase in the number of \ac{VUE}
pairs increases the inter-\ac{VUE} interference, and thus, the
per-user secrecy rate decreases. However, when $v=100$ km/h, the
distance between two neighbouring \ac{VUE} pairs is more than the
coverage range of $100$ m (provided that each next vehicle passes
through a certain point after $5$ s). Hence, the inter-\ac{VUE}
interference is negligible and the per-user secrecy rate almost remains
the same. 

\vspace{-4mm}
\section{Conclusion}
\vspace{-1mm}
In this paper, we proposed two methods for maximizing the sum secrecy rate of all \ac{VUE} pairs in the presence
of \acp{CUE}. Specifically, we have considered the joint optimization of bandwidth reuse coefficients and power control variables and solved
the problem first through the \ac{SCA} method, which is a well-known optimization technique. Later, to reduce the complexity of the \ac{SCA} method, we proposed another first-order low-complexity \ac{FISTA} method. We have showcased through numerical analysis that the proposed \ac{SCA} and \ac{FISTA} methods exhibit a trade-off in terms of convergence and computational complexity. While \ac{SCA} provides better performance, \ac{FISTA} has significantly reduced runtime. Moreover, our findings suggest that when the speed of vehicles is low, the reduced distance between consecutive vehicles within the coverage range results in a decrease in the per-user secrecy rate for pairs of \ac{VUE}. \textcolor{black}{For future work,  it is always worth exploring a more complex system model such as multiuser MIMO with imperfect CSI.}

\section*{Acknowledgment}

\vspace{-2mm}Funding for this research is provided by the European
Union's Horizon Europe research and innovation program through the
Marie Sklodowska-Curie SE grant under the agreement RE-ROUTE No 101086343.

\vspace{-4mm}\bibliographystyle{ieeetr}
\bibliography{IEEEabrv,references}

\end{document}